\documentclass[a4paper,11pt]{article}
\pdfoutput=1 

\usepackage{jcappub} 

\usepackage[T1]{fontenc} 
\usepackage[normalem]{ulem}
\usepackage{amsmath}
\usepackage{amssymb}
\usepackage{geometry} 
\usepackage{booktabs} 
\usepackage{array} 
\usepackage{paralist} 
\usepackage{verbatim} 
\usepackage{graphicx}
\usepackage[backend=biber,style=ieee]{biblatex}
\usepackage[utf8]{inputenc}

\usepackage{wasysym} 
\usepackage{subfigure} 
\usepackage{enumitem}

\title{\boldmath 
Searching for Dark Matter in the Galactic Halo with a Wide Field of View TeV Gamma-ray Observatory in the Southern Hemisphere}

\author[a,1]{Aion Viana, \note{Corresponding author.}}
\author[b]{Harm Schoorlemmer,}
\author[c]{Andrea Albert,}
\author[a]{Vitor de Souza,}
\author[c]{J. Patrick Harding,}
\author[b]{Jim Hinton}

\affiliation[a]{Instituto de F\'isica de S\~ao Carlos, Universidade de S\~ao Paulo, Av. Trabalhador S\~ao-carlense 400, S\~ao Carlos, Brasil.}
\affiliation[b]{Max-Planck Institut f\"ur Kernphysik,\\Saupfercheckweg 1, 69117, Heidelberg, Germany}
\affiliation[c]{Los Alamos National Laboratory}

\emailAdd{aion.viana@ifsc.usp.br}
\emailAdd{harmscho@mpi-hd.mpg.de}
\emailAdd{vitor@ifsc.usp.br}
\emailAdd{jim.hinton@mpi-hd.mpg.de}
\emailAdd{amalbert@lanl.gov}
\emailAdd{jpharding@lanl.gov}

\abstract{Despite mounting evidence that dark matter (DM) exists in the Universe, its fundamental nature remains unknown. We present sensitivity estimates to detect DM particles with a future very-high-energy ($\gtrsim$ TeV) wide field-of-view gamma-ray observatory in the Southern Hemisphere. This observatory would search for gamma rays from the annihilation or decay of DM particles in the Galactic halo. With a wide field of view, both the Galactic Center and a large fraction of the Galactic halo will be detectable with unprecedented sensitivity to DM in the mass range of $\sim$500 GeV to $\sim$2 PeV. 
These results, combined with those from other present and future gamma-ray observatories, will likely probe the thermal relic annihilation cross section of Weakly Interacting Massive Particles for all masses from $\sim$80 TeV down to the GeV range in most annihilation channels. }

\begin{document}
\maketitle
\flushbottom

\section{Introduction} 

The nature of Dark Matter (DM) is one of the most fundamental open questions in physics. There are many observations pointing to the existence of DM -- from galaxy rotation curves~\cite{RotationCurves}, galaxy cluster dynamics~\cite{BulletCluster}, the cosmic microwave background fluctuations~\cite{Planck2013}, and others. Despite several searches, no DM particle signal has ever been detected. 

The most promising candidates are called Weakly Interacting Massive Particles (WIMPs). These are non-baryonic particles with masses in the GeV-TeV range and weak-scale interaction strength. WIMPs are stable, electrically neutral, and if they were in thermal equilibrium in the early Universe, they could produce the observed abundance of DM today.
Other models have expanded the potential DM particle mass range to include PeV masses and stronger interactions or even decaying DM (\textit{e.g.} dark glueballs~\cite{Acharya:2017szw,Boddy:2014yra,Cohen:2016uyg,Faraggi:2000pv,Forestell:2016qhc,Halverson:2016nfq,Soni:2017nlm} and hidden sector DM~\cite{Berlin:2016vnh,Berlin:2016gtr}).

Human-made collider experiments~\cite{Boveia:2018yeb} and large passive calorimeters~\cite{Liu:2017drf} are probing the DM mass range up to hundreds of GeV. However, if the DM is composed of particles with a mass well above the TeV scale, the only near-future discovery possibilities may be astrophysical. This is because the current generation of colliders do not reach sufficiently high energies to produce $>$TeV DM, and the naturally-produced flux of DM above the TeV mass scale is not likely to be reachable with the current target masses and technology of the direct-detection experiments.

Via the massive DM halos associated to astrophysical objects on large scales, and the high-energy reach of astrophysical experiments, DM particle masses much greater than 1~TeV can be probed through their annihilation or decay products. Several experiments are currently utilised to search for such signatures~\cite{dSphHAWC,Abeysekara:2017jxs,HAWC:2018eaa,Abdallah:2016ygi,Abdalla:2018mve,Abdallah:2018qtu,Ahnen:2016qkx,Acciari:2018sjn,Archambault:2017wyh,Fermi-LAT:2016uux} and several more are planned~\cite{Doro:2012xx,CTA_ScienceTDR,He:2019dya}.
Although unambiguous evidence of particle DM is still absent, the allowed parameter space is increasingly constrained by such searches.

Gamma-ray observatories are particularly promising for the DM search at masses beyond 1~TeV, with high sensitivity and sufficient angular resolution obtainable. 
Given its large DM content and relative proximity, the Galactic Center (GC) region is expected to be the brightest source of gamma rays from DM annihilation or decay in the sky by several orders of magnitude. Even considering possible signal contamination from other astrophysical sources, it is one of the most promising targets to detect the presence of new massive DM particles. 

A TeV gamma-ray observatory with a very wide field of view (FOV) is a promising possibility in the search for very extended low-surface brightness emission associated with the DM halos of our own galaxy.
Several observatories of this kind are operational in the Northern Hemisphere (in particular LHAASO \cite{Bai:2019khm} and HAWC \cite{HAWC_CRAB}) using arrays of particle detectors located at high-elevation sites. A similar facility located in the Southern Hemisphere would be highly sensitive to DM gamma-ray signals from the GC region, in particular because the GC would transit close to zenith in relation to such a detector. An observatory of this type has been proposed for construction in South America and recently a collaboration was formed to start the research and development phase under the name \textit{Southern Wide field-of-view Gamma-ray Observatory} (SWGO\footnote{\url{www.swgo.org}})\cite{Schoorlemmer2019}. Sensitivity at the highest energies ($>10$ TeV) is only achievable with a large detection area ($>50000$\,m$^2$) and a long exposure. This can be achieved for the entire observable sky by a passive, wide FOV detector of gamma rays, complementing searches with pointed systems such as Imaging Atmospheric Cherenkov Telescopes (IACTs). This next generation observatory will have unprecedented sensitivity in the multi-TeV energy scale, large FOV (45$^\circ$), daily exposure of the GC, and sufficient angular resolution ($<0.5^\circ$). The large FOV allows for the observation of a large fraction of the Galactic DM halo, which reduces the dependence of the search on the DM radial profile. It would, therefore, be sensitive to several different DM profile models (\textit{e.g.} cuspy or cored). In addition, above the TeV scale, background from other astrophysical sources is expected to be very modest away from the Galactic Plane, allowing the detection of very faint DM signals.

We investigate in detail the potential of an experiment such as SWGO to measure a DM signal. Based on a ``straw man" detector design of SWGO, we derive sensitivity limits to the annihilation and decay of DM particles in the Galactic halo. We show that, together with other contemporaneous gamma-ray observatories, SWGO would allow the detection of DM particles with a thermal relic annihilation cross-section ($\sim 3 \times 10^{-26}$ cm$^3$ s$^{-1}$~\cite{Beacom:2006tt}) for masses up to 100\,TeV. We also investigate the impact of different DM halo density profiles on the sensitivity limits, as well as how electroweak (EW) radiative corrections improve the sensitivity of SWGO to DM annihilation/decay signals.

\section{A Straw Man Design for a Southern Gamma-ray Survey Observatory}

\subsection{Instrumental Context}
A wide FOV very-high-energy gamma-ray detector in the Southern Hemisphere is now being discussed~\cite{SGSO_WP}. It would complement existing and future Northern Hemisphere instruments like HAWC~\cite{HAWC_CRAB} and LHAASO~\cite{Bai:2019khm}. It would also complement the CTA Observatory, which will have both Northern and Southern Hemisphere arrays. A Southern Hemisphere location is essential for targeting the inner Galactic halo, which is the most promising DM target, \textit{i.e.} the one with the largest astrophysical $J$-factor~\ref{eq:Jfactors}. 

The Fermi-LAT instrument has very limited detection capabilities for WIMP masses beyond $\sim$100~GeV due to its limited collection area. CTA will have excellent performance up to at least a few TeV from deep, targeted observations of the region around the GC~\cite{CTA_ScienceTDR}. However, CTA limited FOV does imply performance limitations in the case of a rather cored central DM density profile, as well as for DM decay. At the highest energies, beyond $\sim$10~TeV in gamma rays, the performance of SWGO is expected to surpass that of CTA for steady point-like sources after a few years of operation~\cite{SGSO_WP}. For emission extended on degree scales, the cross-over point in sensitivity occurs at lower energies, due to the worsening of CTA sensitivity for extended sources. The combination of a very wide FOV and excellent performance at multi-TeV energies make a SWGO-like detector an exciting prospect for heavy WIMP detection from the Galactic halo.

\subsection{Simulations Results}
The detector model of a potential SWGO-like observatory design is used to assess the sensitivity to DM searches. For this purpose, publicly-available\footnote{\url{https://github.com/harmscho/SGSOSensitivity}} instrument response functions were used, which have been produced for the science case studies presented in Ref. \cite{SGSO_WP}. This observatory design can be regarded as a scaled-up version of the current generation of its type, like High Altitude Water Cherenkov gamma-ray observatory (HAWC) \cite{HAWC_CRAB}. In our calculation a latitude of 25$^{\circ}$S is assumed, giving close to optimal exposure to the GC. Currently several sites in the range from 10$^{\circ}$S to 30$^{\circ}$S latitude are under consideration for the location of SWGO.

While the response of a single detector unit has been kept roughly similar to HAWC, the size of the array, the ground coverage, and elevation have been significantly increased with respect to HAWC. The ground coverage was increased to 80\%, which is between the ground coverage of HAWC (57\%) and LHAASO (~100\%). The elevation is chosen at 5\,km altitude, since there are several site candidates with an elevation close to this altitude under consideration by the SWGO collaboration.

To estimate realistic instrument response functions, published performance figures from HAWC \cite{HAWC_CRAB} (like angular resolution and gamma and hadron cut passing rate) are used as a baseline. To these we relate the instrument response functions of the straw man design by assuming that the performance stays the same when the shower deposits the same amount of energy on the array. The energy deposited on the array is estimated for a HAWC-like detector and the straw man design using a toy detector together with air shower simulations generated by CORSIKA \cite{corsika} (more details can be found in \cite{SGSO_WP}). With the improved array parameters for the straw man design, a similar performance will be reached at a lower gamma-ray energy than HAWC.
\begin{figure}[ht]
  \begin{center}	
    \includegraphics[width=0.49\linewidth]{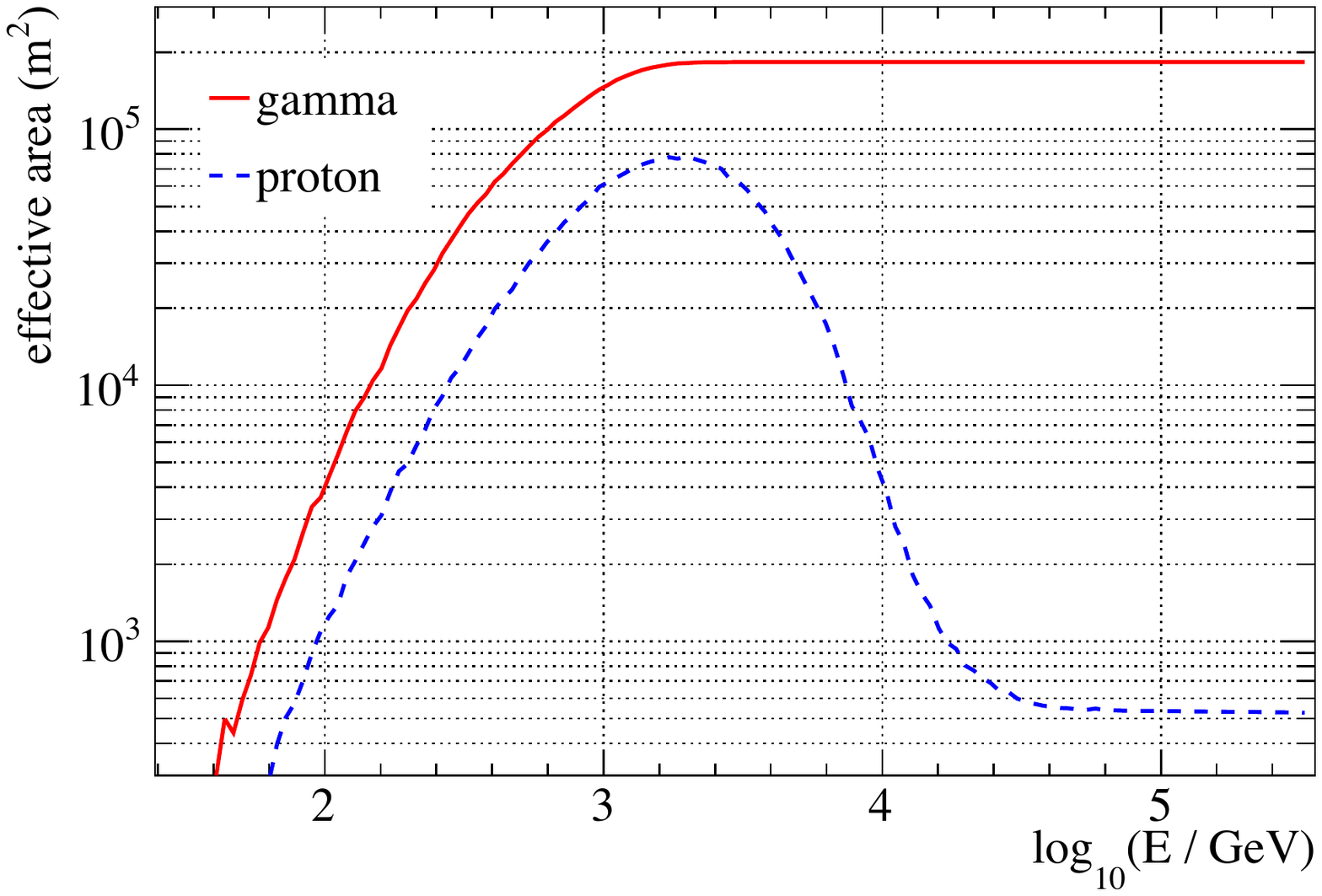}
    \includegraphics[width=0.49\linewidth]{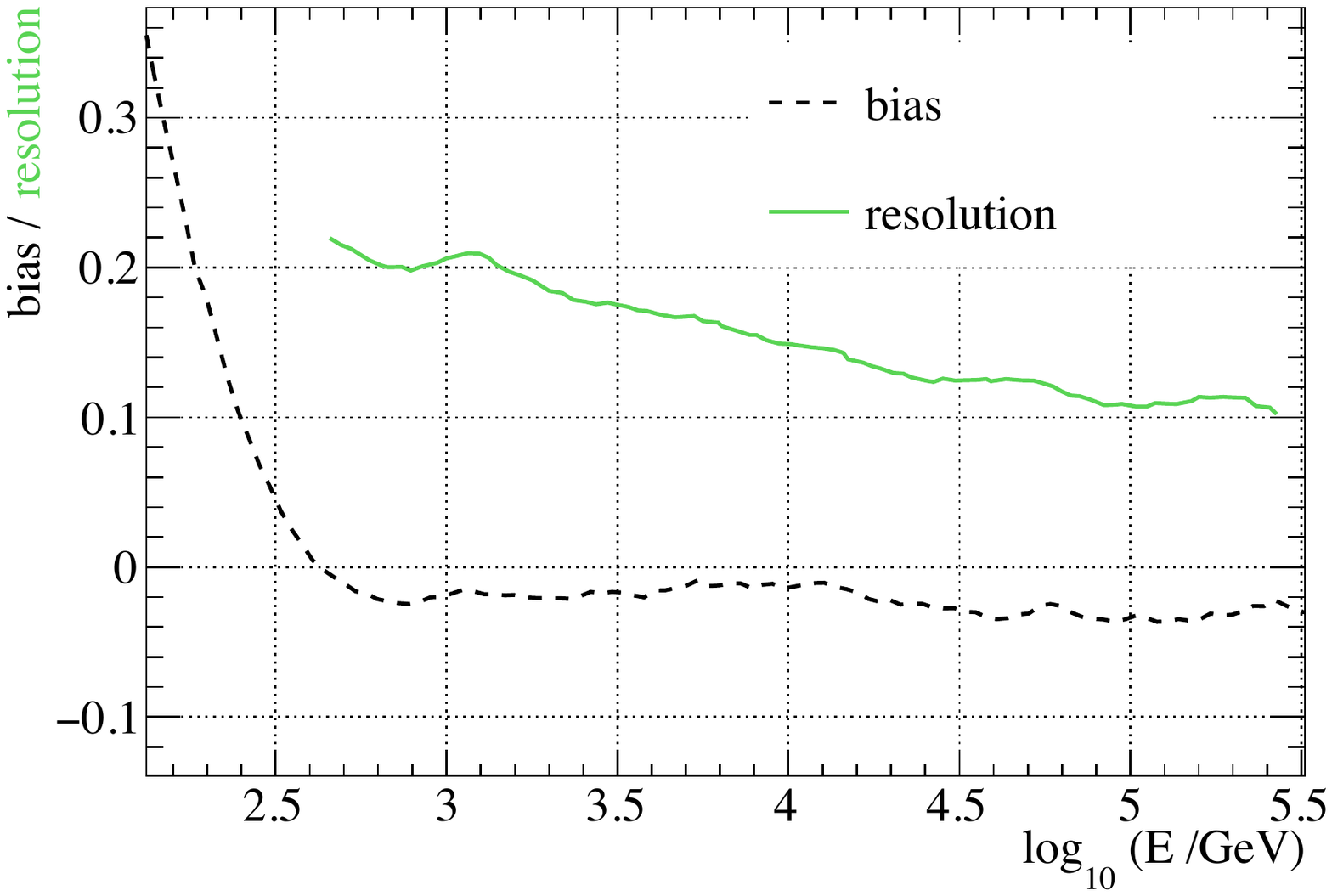}
   \end{center}
  \caption{{\em Left:} Effective area for a source at $20^{\circ}$ from zenith as a function of energy of the primary particle (gamma ray or proton) after applying gamma-hadron separation and trigger multiplicity cuts. {\em Right:} Energy bias and resolution, where bias is defined as mean value of $\Delta = (\log_{10}{\hat{E}} - \log_{10}{E}$) (with reconstructed energy  $\hat{E}$ and true energy $E$), while the resolution in taken as the root mean square of $\Delta$. In both the figures the x-axis gives the true energy of the primary particle.}
  \label{fig:SGSOperfomance}
\end{figure}
For the study of DM signals from the inner Galactic halo, the DM mass regime beyond the energy range of CTA provides a unique opportunity for an air shower array.  To probe this regime, a sufficiently larger effective area is of key importance. The effective area of the straw man design, after the application of gamma-ray selection cuts, is shown in Figure \ref{fig:SGSOperfomance} for simulated gamma rays and proton primary particles. The ratio of the proton and gamma-ray effective areas indicates the efficiency of the cosmic-ray induced background rejection, which is better than 99.9\% above a gamma-ray energy of $\sim$15\,TeV.  At $\sim$15\,TeV, the background rejection is similar to  CTA-South\footnote{\url{https://www.cta-observatory.org/science/cta-performance/}}, while at higher (lower) gamma-ray energy the background rejection is better (worse) than expected for CTA-South. The proton rejection efficiency has been scaled from the performance of HAWC\cite{SGSO_WP} and saturates at the HAWC background rejection efficiency at the highest energy. This is conservative approach as no intrinsic improvement over HAWC was assumed. However, in the final design of SWGO significant improvements in background rejection can be expected.
 
Another important factor is the energy migration matrix (relation between true and reconstructed gamma-ray energy), which is summarized by the bias and resolution of the difference between true $E$ and reconstructed $\hat{E}$ energy, shown in the right panel of Figure \ref{fig:SGSOperfomance}. The difference in true  and reconstructed  gamma-ray energy follows roughly Gaussian distribution in log-space $\Delta =(\log_{10}{\hat{E}} - \log_{10}{E})$ above $\hat{E}\approx$500\,GeV. However, for the calculations presented here the full energy migration matrix was used to also account for the larger bias and asymmetric shape that occur at gamma-ray energy below 500\,GeV.  Since the root mean square is not a good approximation of the width of the distribution below this energy, the resolution curve is only drawn above this energy. 
The obtainable resolution becomes especially relevant in the case where there are pronounced spectral features in the gamma-ray DM annihilation spectrum that would provide a unique signature. The expected energy resolution is below 40\% above a gamma-ray energy of 10\,TeV. Since a very simplistic energy assignment was used, this should be considered as an upper limit on the energy resolution of a future observatory. 

The angular resolution also needs to be sufficient to resolve the angular scales of the emission, but with an anticipated angular resolution of less than 0.3$^\circ$ above 10 TeV this should not be a limiting factor. 




\section{Gamma-ray Fluxes from Dark Matter towards the Galactic Center} 

\subsection{Annihilation and Decay of Dark Matter Particles}
\hfill \break
The prompt gamma-ray flux from the annihilations (${\rm d}\Phi_{\rm Ann}/{\rm d} E_{\gamma}$) and decays (${\rm d}\Phi_{\rm Dec}/{\rm d} E_{\gamma}$) of DM particles of mass $M_{\rm DM}$ in a DM halo are given by a particle physics term (left parenthesis) times an astrophysical term (right parenthesis):
\begin{equation}
\label{eq:dm_flux_ann}
\frac{{\rm d}\Phi_{\rm Ann}(\Delta\Omega,E_{\gamma})}{{\rm d} E_{\gamma}}\,= \left(\frac12 \frac{1}{4\pi}\, \frac{\langle \sigma v \rangle}{M_{\rm DM}^2}
\frac{{\rm d} N}{{\rm d}E_\gamma} \right) \,\times\, \left(J(\Delta\Omega)\right) \, ,
\end{equation}
and  
\begin{equation}
\label{eq:dm_flux_dec}
\frac{{\rm d}\Phi_{\rm Dec}(\Delta\Omega,E_{\gamma})}{{\rm d} E_{\gamma}}\,= \left(\frac{1}{4\pi}\, \frac{1}{\tau_{\rm DM} M_{\rm DM}}
\frac{{\rm d} N}{{\rm d}E_\gamma} \right) \,\times\, \left(D(\Delta\Omega)\right) \, .
\end{equation}
The astrophysical factors, also called $J$-factor for annihilations and $D$-factor for decays, are integrated over a given region of interest (ROI) of solid angle size $\Delta \Omega$ along the line of sight (l.o.s.). They are defined as
\begin{equation}
J(\Delta\Omega) = \int_{\Delta \Omega}  \int_{\rm l.o.s.} {\rm d}\Omega \, {\rm d} s \ \rho_{\rm DM}^2[r(s,\Omega)] \, ,
\label{eq:Jfactors}
\end{equation}
\begin{equation}
D(\Delta\Omega) = \int_{\Delta \Omega}  \int_{\rm l.o.s.} {\rm d}\Omega \, {\rm d} s \ \rho_{\rm DM}[r(s,\Omega)] \, ,
\label{eq:Dfactors}
\end{equation}
where $\rho_{\rm DM}$ is the DM density distribution. 

The particle physics term contains  $M_{\rm DM}$, the velocity-weighted annihilation cross section $\langle \sigma v\rangle$, DM lifetime $\tau_{\rm DM}$, and the differential spectrum of gamma rays in a specific annihilation or decay channel ${\rm d} N/{\rm d} E_{\gamma}$. We take a model-independent
 approach by considering DM particles annihilating/decaying into different, single channels with a 100\% branching ratio. As representatives of different types of Standard Model (SM) particles, we compute our limits for annihilation/decay into pairs of gauge bosons, $W^+ W^-$, quarks, $b\bar{b}$, and leptons, $\tau^+\tau^-$. In the cases where positrons and electrons are produced in the final states, an additional contribution to the gamma-ray flux can come from Inverse Compton (IC) up-scattering of ambient photons, such as those of the cosmic microwave background (CMB). However, this additional contribution is sub-dominant at TeV energies. Thus, here we only consider the prompt gamma-ray emission, which leads to a slightly conservative estimate of the SWGO sensitivity to a DM signal. In Figure \ref{fig:profile_comparison} the gamma-ray energy distributions are compared for the three annihilation channels with a DM of 10\,TeV.  

Usually, searches for DM annihilation focus on particle masses below a few hundred TeV. Within this mass range, most models of DM particles will produce the DM thermal relic abundance without being in violation of the unitarity bound~\cite{Griest:1989wd,Hui:2001wy,Beacom:2006tt} (for some exceptions, see for \textit{e.g.} ref.~\cite{Berlin:2016vnh,Berlin:2016gtr}). Thus, for the case of DM annihilation, we limit ourselves to masses between 500 GeV and 100 TeV, and we extract the final state gamma-ray spectra from the PPPC 4 DM ID~\cite{Cirelli:2010xx}. There is, however, no theoretical limit to DM particle masses for the case of decaying DM, so we extend our limits to DM particles as massive as $\sim$2 PeV. In this case, we produced ourselves the final-state gamma-ray spectra using the PYTHIA 8.219 software~\cite{Sjostrand:2006za, Sjostrand:2014zea} with electroweak corrections enabled~\cite{Christiansen:2014kba}. 


\subsection{Galactic halo density profiles and regions of interest}

\begin{figure}[h!]
\includegraphics[width=0.45\linewidth]{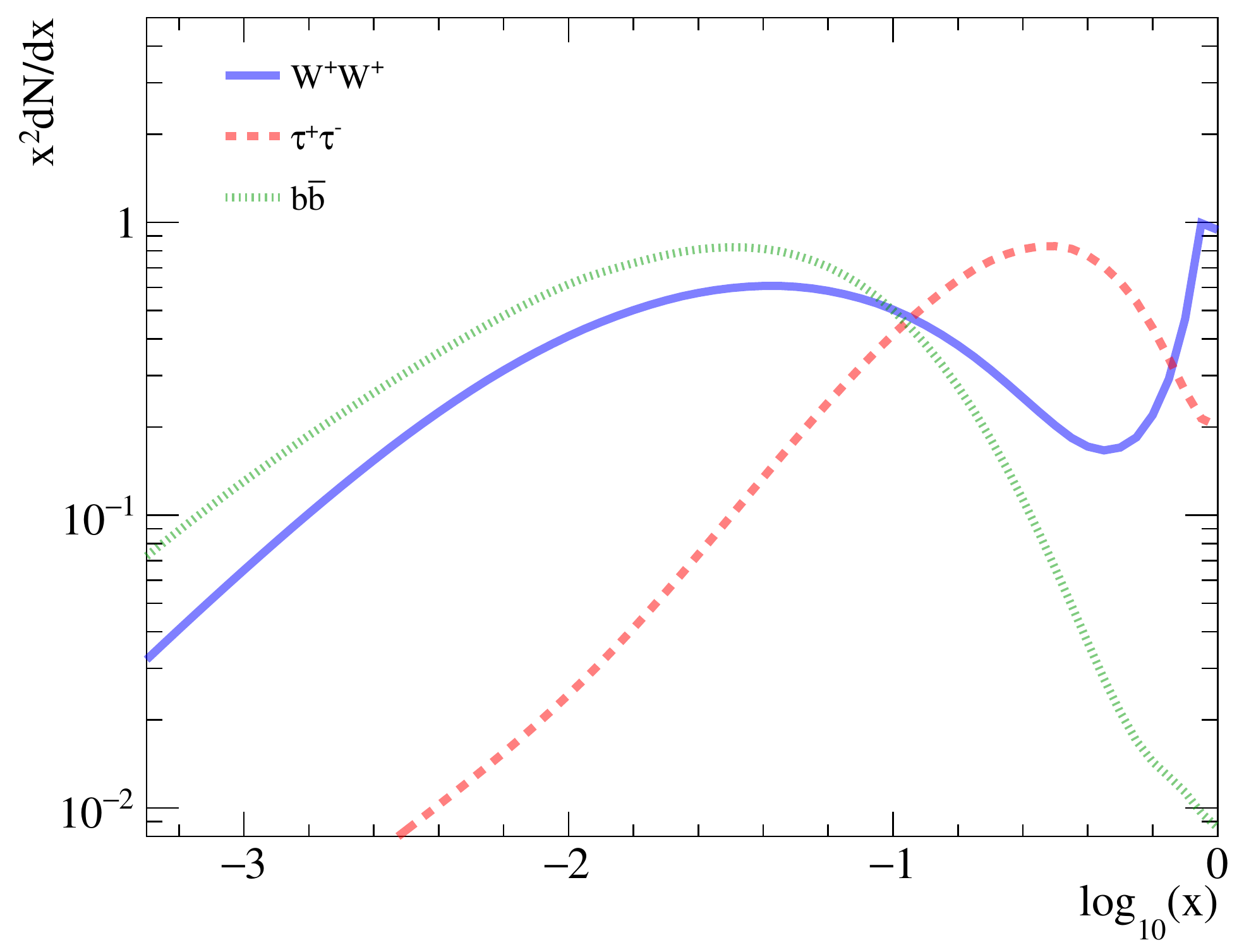}
\includegraphics[width=0.45\linewidth]{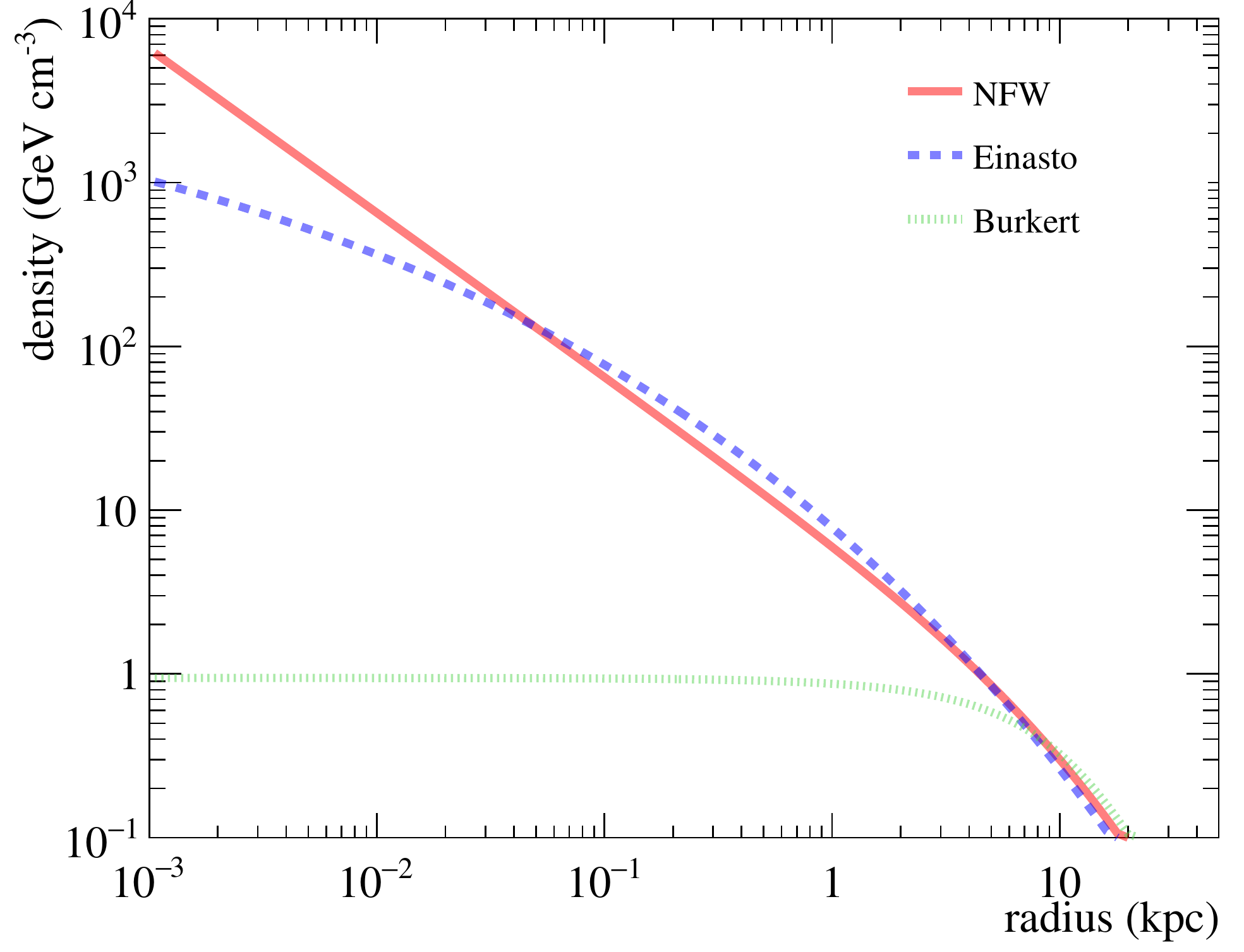}
\caption{\emph{Left:} Comparison of the gamma-ray energy density distribution of three annihilation channels for $M_{\text{DM}} = 10$\,TeV (and $x = E_{\gamma}/M_{\text{DM}}$ ) \cite{Cirelli:2010xx}. \emph{Right:} Behavior of three physically-motivated DM density profiles as a function of radial distance from the center.}
\label{fig:profile_comparison}	
\end{figure}

One of the main difficulties when searching for DM signals from the Milky Way is that the DM density distribution of the Galactic halo is poorly constrained. As shown in Fig. \ref{fig:profile_comparison}, the expected DM density varies greatly between possibly functional forms, with the Einasto and NFW ("cuspy")  profiles peaking sharply, and the Burkert ("cored") profile levelling off. This creates substantial uncertainty in the $J$ and $D$-factors and therefore on the corresponding sensitivity, especially for searches close to the center of the halo. 

Thus, in order to estimate the predicted DM flux, it is important to consider different classes of halos. Here two models are assumed: a peaked Einasto profile~\cite{Pieri:2009je} and a cored Burkert profile~\cite{Burkert:1995yz}, parametrised as

\begin{equation}
\rho_{E} (r) = \rho_0 \exp \left \{ \frac{-2}{\alpha} \left[ \left( \frac{r}{r_s} \right)^{\alpha} -1 \right] \right \} \, 
\end{equation}
and
\begin{equation}
\rho_B(r) = \frac{\rho_c r_c^3}{(r+r_c)(r^2 + r_c^2)} \, ,
\label{eq:burk}
\end{equation}
respectively. Here $r_s$ and $\rho_0$ are the radius and density at which the logarithmic slope of the density is -2, respectively, $\alpha$ is a parameter describing the degree of curvature of the profile, $\rho_c$ is the central density, and $r_c$ the core radius. We take explicitly $r_s = 20$ kpc, $\alpha = 0.17$~\cite{Pieri:2009je} and $r_c = 12.67$ kpc ~\cite{Cirelli:2010xx}. $\rho_0$ and $\rho_c$ are chosen so that the local DM density measures $\rho_{\rm DM} (r_{\odot}) = 0.39$ GeV/cm$^3$, where $r_{\odot}$ is the distance from the Sun to the GC~\cite{Abramowski:2011hc, Catena:2009mf}. 

We focus our searches for DM signals to the inner 10$^{\circ}$ of the Galaxy. The spatial ROIs are defined as circular concentric regions of 0.2$^{\circ}$ width each, centered at the GC, excluding a $\pm$0.3$^{\circ}$ band in Galactic latitude to avoid the standard astrophysical background in the TeV energy range. The latter includes the diffuse emission from the Galactic Center Ridge~\cite{Aharonian:2006au,Abramowski:2016mir}, the GC central gamma-ray source~\cite{Aharonian:2004wa}, the composite supernova remnant G 0.9+0.1~\cite{Aharonian:2005br}, and a few other weaker sources~\cite{Aharonian:2005kn,Abramowski:2011yya,Smith:2015pya,Ahnen:2016crz,Abdalla:2017xja}. If other sources are found beyond $\pm$0.3$^{\circ}$ latitude, extra exclusion regions may be added around them, hence decreasing the available region for DM searches accordingly. Alternatively, if the origin of these sources is known, they may be modelled and subtracted as background. In any case, if the size of the sources is not too large when compared to the ROIs used here for DM searches, their impact to the sensitivity estimates should not be very significant. In Table~\ref{tab:JDfactors} we present the solid angle sizes $\Delta \Omega_i$, $J$-factors and $D$-factors calculated for these ROIs. 

\begin{table}[htb]
	\small
	\centering
	\begin{tabular}{|c|c|cc|cc|}
		\hline\
		 $i$-th ROI & Solid Angle & \multicolumn{2}{c|}{$J(\Delta\Omega_i)$ [10$^{19}$ GeV$^2$ cm$^{-5}$]}  & \multicolumn{2}{c|}{$D(\Delta\Omega_i)$ [10$^{19}$ GeV cm$^{-2}$]} \\
	    $\Delta \theta_i = [\theta_{\rm min},\theta_{\rm max}]$	& $\Delta \Omega_i$ [10$^{-4}$ sr] & Einasto & Burkert & Einasto & Burkert   \\ \hline
		$\Delta \theta_1 = [0.3^{\circ},0.5^{\circ}]$ & 0.68  & 75.78  & 0.17 & 1.91 & 0.31  \\ 
		$\Delta \theta_2 = [0.5^{\circ},0.7^{\circ}]$ & 1.53  & 129.11  & 0.38 & 4.06 & 0.70  \\ 
		$\Delta \theta_3 = [0.7^{\circ},0.9^{\circ}]$ &  2.31 & 154.19  & 0.58 & 5.83 & 1.06  \\ 
		$\Delta \theta_4 = [0.9^{\circ},1.1^{\circ}]$ & 3.08  & 168.57  & 0.78 & 7.43 & 1.41  \\ 
		\vdots & \vdots  & \vdots  & \vdots & \vdots & \vdots  \\ 
		$\Delta \theta_{45} = [9.1^{\circ},9.3^{\circ}]$ &  34.33 & 110.84  & 8.10 & 36.86 & 15.33  \\
		$\Delta \theta_{46} = [9.3^{\circ},9.5^{\circ}]$ &  35.09 & 109.28  & 8.25 & 37.23 & 15.65  \\
		$\Delta \theta_{47} = [9.5^{\circ},9.7^{\circ}]$ &  35.84 & 107.76  & 8.41 & 37.59 & 15.97  \\
		$\Delta \theta_{48} = [9.7^{\circ},9.9^{\circ}]$ & 36.60  & 106.27  & 8.46 & 37.94 & 16.29  \\ \hline
		$\Delta \theta_{\rm total} = [0.3^{\circ},9.9^{\circ}]$ & 899.5  & 7032.9  & 218.3 & 1190.0 & 405.7  \\
		\hline
	\end{tabular}
	\caption{Definitions of the ROIs with their corresponding inner ($\theta_{\rm min}$) and outer ($\theta_{\rm max}$) radii, the solid angle of each ROI, and values of $J$-factors and $D$-factors calculated for both Einasto and Burkert profiles. Only the first and last 4 ROIs out of the 48 are presented here.  \label{tab:JDfactors} }
\end{table}

\subsection{Analysis methodology}

The sensitivity of SWGO to DM annihilation/decay can be found by comparing the number of observable gamma rays with the expected background. This residual background is essentially composed of protons that are misidentified as gamma rays (see left panel of Fig~\ref{fig:SGSOperfomance} for the simulation results, and ref~\cite{SGSO_WP} for more details). The statistical tool used to derive limits is a 2D (energy and space) joint-likelihood method, where the comparisons between DM and background fluxes are performed in different energy and spatial intervals (or bins)~\cite{Abdallah:2016ygi}. Hereafter, we divide the energy range between 100 GeV and 100 TeV into 40 logarithmically-spaced bins.  The number of observable gamma-ray events by a detector is computed by folding the considered gamma-ray flux with the instrument response functions. The expected signal in a spatial ROI $i$ and energy bin $j$ is given by

\begin{equation}
S_{ij} = T_{\rm obs} \int_{\Delta E_j} d\hat{E} \int_0^{\infty} dE \frac{{\rm d}\Phi_{\gamma}(\Delta \Omega_i, E)}{{\rm d} E} \times A_{\rm eff}(E) \times {\rm PDF}(E,\hat{E})
\end{equation}  
where $T_{\rm obs}$ is the observation time, $E$ is the true primary energy, $A_{\rm eff}$ is the effective collection area as function of the true energy (Fig. \ref{fig:SGSOperfomance} right), and ${\rm PDF}(E,\hat{E})$ is the representation of the energy resolution as the probability density function $P(\hat{E}|E)$, of observing an event at the reconstructed energy $\hat{E}$ for a given true energy $E$.

Assuming that the number of detected events follows a Poisson distribution, the likelihood functions are calculated in each individual bin and combined into a joint-likelihood function:

\begin{equation}
\mathcal{L} (M_{\rm DM},\langle \sigma v \rangle \lor \tau_{\rm DM}) = \prod_{ij} \mathcal{L}_{ij} \,,
\end{equation}

with

\begin{equation}
\mathcal{L}_{ij} = \frac{(B_{ij} + S_{ij})^{N_{ij}} \exp(-B_{ij} + S_{ij})}{N_{ij}!}\,.
\end{equation}

Here $B_{ij}$ is the number of background observed counts, and $N_{ij}$ is the total number of observed counts. Constraints on $\langle \sigma v \rangle$ (or $\tau_{\rm DM}$) are obtained from the log-likelihood ratio test statistic given by $TS = - \ln(\mathcal{L}_0(M_{\rm DM},\langle \sigma v \rangle \lor \tau_{\rm DM})/\mathcal{L}_{\rm max}(M_{\rm DM},\langle \sigma v \rangle) \lor \tau_{\rm DM})$, where $\mathcal{L}_0$ is the null hypothesis (no DM model) likelihood and $\mathcal{L}_{\rm max}$ is the alternative hypothesis (with DM model) likelihood, evaluated at the value of the cross-section (or decay time) which maximizes the likelihood. Values of $\langle \sigma v \rangle$ (or $\tau_{\rm DM}$) for which TS is higher than 2.71 are excluded at 95\% confidence level (C.L.). This method takes full advantage of differences on the energy and spatial distribution between the expected DM signal and the background. For instance, the former is supposed to follow the $J$-factor ($D$-factor), whereas the latter is isotropic on the sky. 

\section{Results}

\subsection{Sensitivity to Dark Matter Annihilation}

\begin{figure}[h!]
	\begin{center}	
		\includegraphics[width=0.49\linewidth]{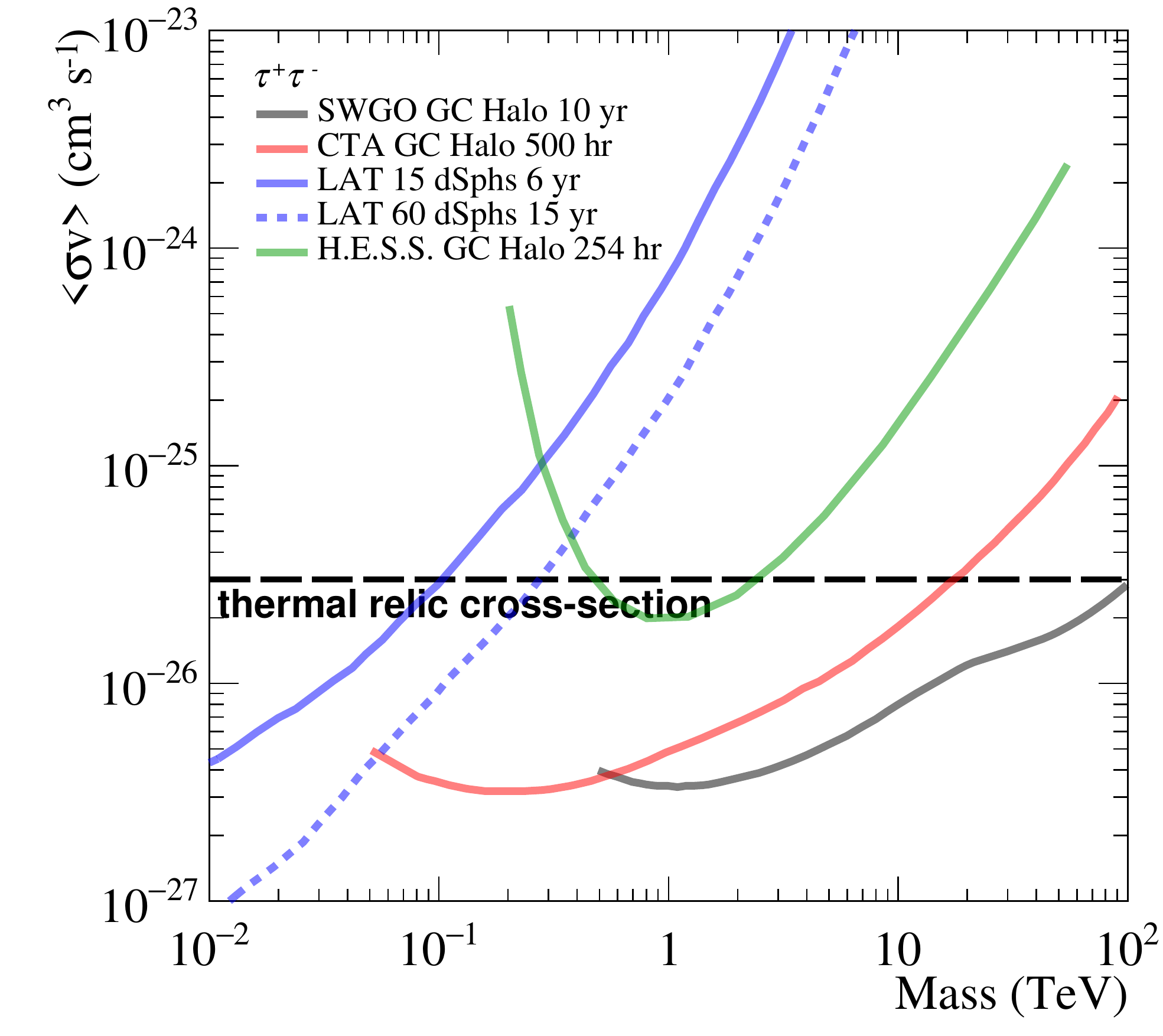}
		\includegraphics[width=0.485\linewidth]{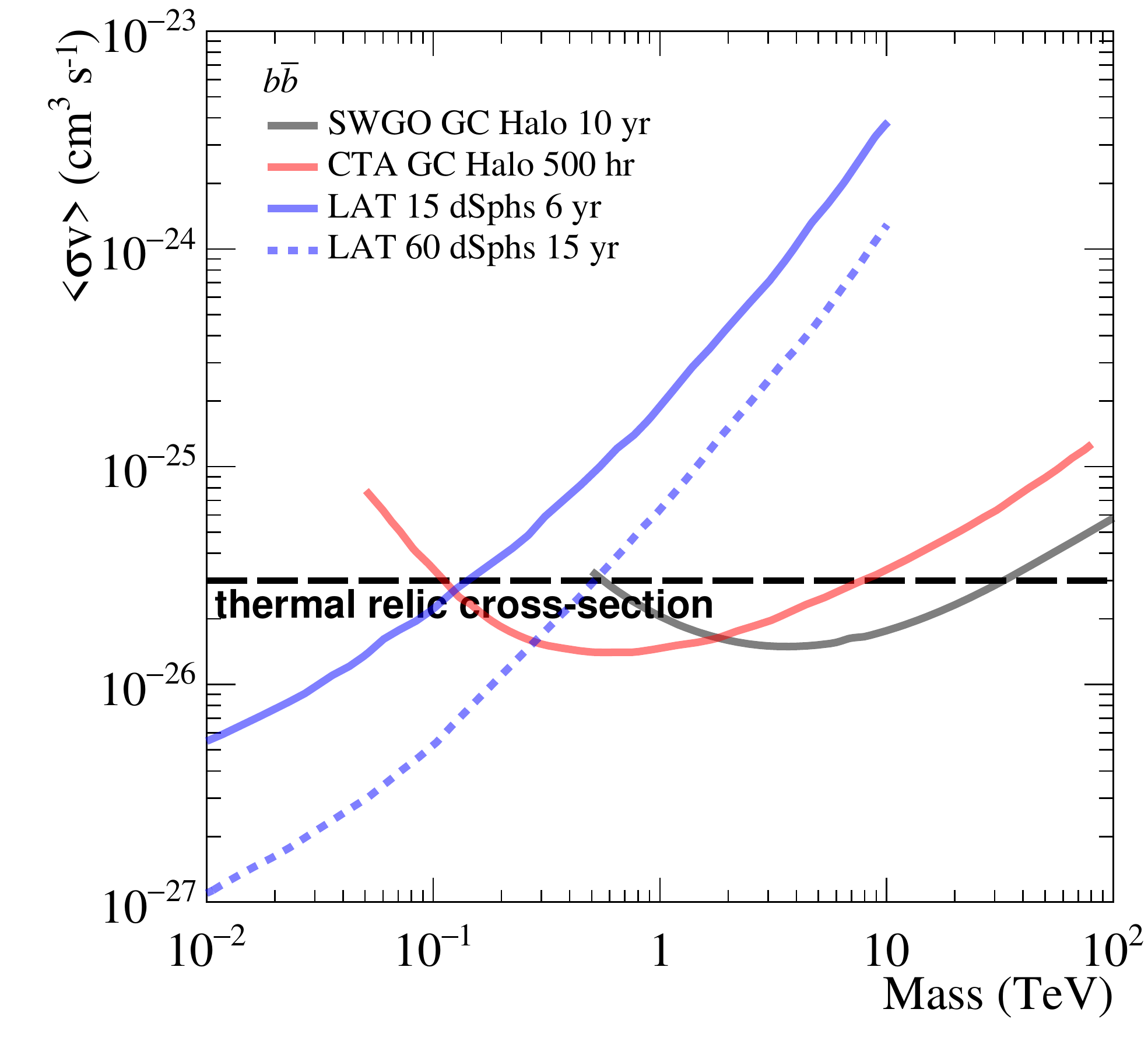}
		\includegraphics[width=0.49\linewidth]{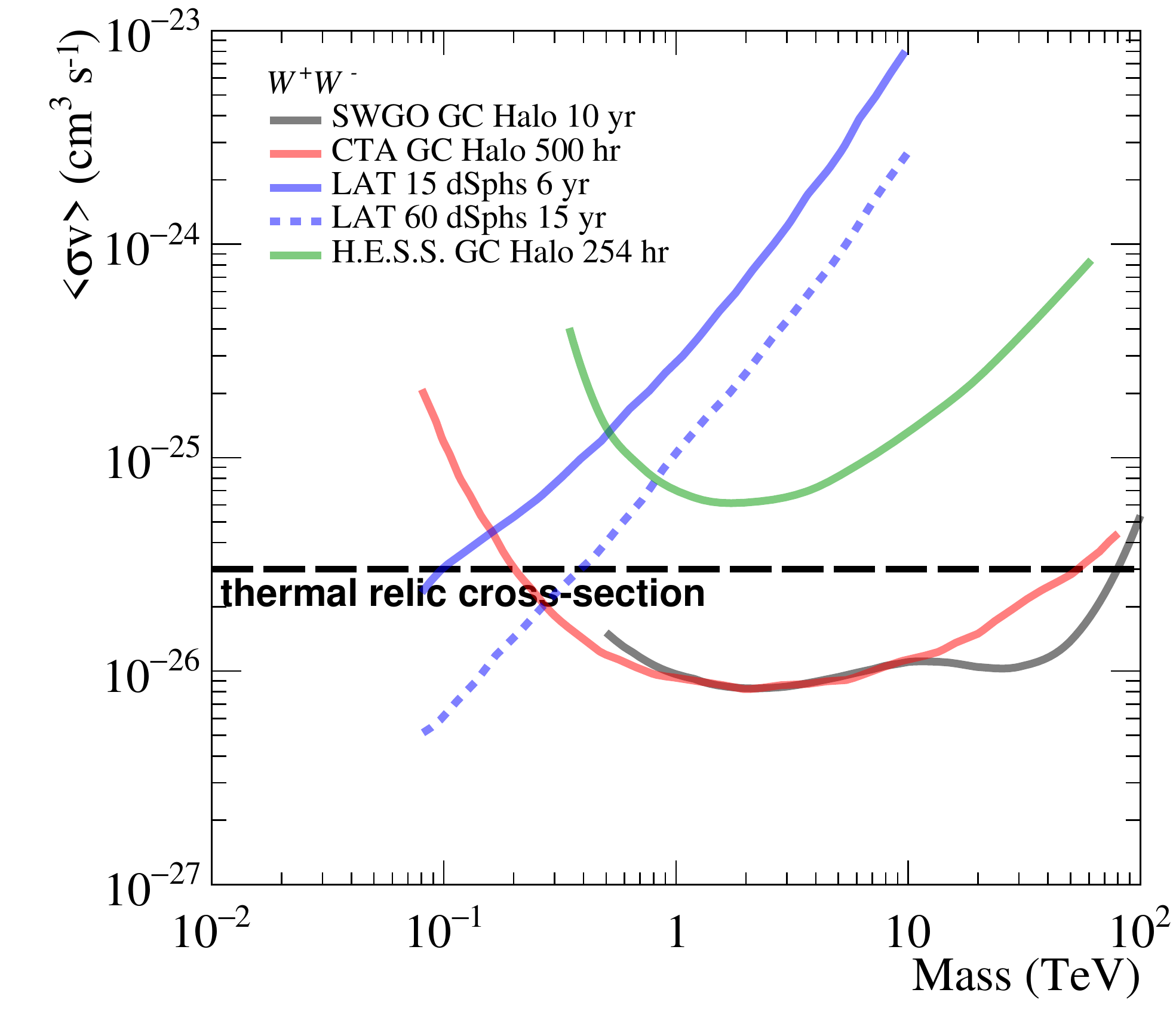}
	\end{center}
	\caption{Expected 95\% C.L. upper limit on the velocity-weighted cross section for DM self-annihilation into $\tau^+ \tau^-$ (top-left), $b\bar{b}$ (top-right) and $W^+ W^-$ (bottom) as a function of $M_{\rm DM}$, for SWGO and CTA~\cite{CTA_ScienceTDR} observations of the GC halo. Current Galactic Center H.E.S.S. limits~\cite{Abdallah:2016ygi} towards the GC halo, and Fermi-LAT limits towards dwarf galaxies (solid blue line) as well as projected sensitivities (dashed blue line) are also plotted~\cite{Fermi-LAT:2016uux}. The nominal value of the thermal-relic cross-section is plotted as well (long-dashed black line).}
	\label{fig:DM_sens}
\end{figure}

Figure~\ref{fig:DM_sens} shows the expected 95\% C.L. upper limits on $\langle \sigma v \rangle$ versus $M_{\rm DM}$ for DM particles annihilating into $W^+ W^-$, $b\bar{b}$ and $\tau^+\tau^-$~\cite{Cirelli:2010xx} assuming an Einasto profile for the GC halo. The sensitivities are shown for 10 years of observations with SWGO. They are compared to the current best limits in the TeV mass range by H.E.S.S.~\cite{Abdallah:2016ygi} and to the future sensitivity of CTA-South~\cite{CTA_ScienceTDR}, with 254 and 500 hours of observation of the inner 1$^{\circ}$ and 5$^{\circ}$ of the Galaxy, respectively, and assuming an Einasto profile in both cases. The current most-stringent Fermi-LAT limits using 15 dwarf spheroidal galaxies (dSphs) are also plotted~\cite{Fermi-LAT:2016uux}, as well as the projected sensitivities assuming a total 15 years of observations and a projected sample of 60 dSphs~\footnote{Here we assumed that the Fermi-LAT $W^+ W^-$ projected sensitivity scales similar to the $b\bar{b}$ sensitivity.}~\cite{Charles:2016pgz}. 


A sensitivity to values of $\langle \sigma v \rangle$ smaller than the nominal thermal-relic cross-section is reachable for SWGO in the mass range of $\sim$500 GeV to $\sim$80 TeV for the $W^+ W^-$ and $\tau^+ \tau^-$ channels, and in the range of $\sim$700 GeV to $\sim$20 TeV for the $b\bar{b}$ channel. SWGO will improve the sensitivity to DM annihilation in the TeV mass range by more than an order of magnitude with respect to the current observatories, such as H.E.S.S.~\cite{Abdallah:2016ygi}. It will also be more sensitive than CTA for all DM particles masses above 700 GeV in the $\tau^+ \tau^-$ channel, and above $\sim$2.5 TeV in the $b\bar{b}$ channel. In the $W^+ W^-$, SWGO will have a similar sensitivity to CTA in the mass range of $\sim$500 GeV to $\sim$20 TeV, and better above 10 TeV. Most importantly, the combined sensitivity of SWGO with Fermi-LAT and CTA will be able to probe a thermal-relic cross-section for all WIMP masses between a few GeV and $\sim 80$ TeV in most annihilation channels ($\lesssim 20$ TeV for $b\bar{b}$).  

\subsection{Sensitivity to Dark Matter Decay}

Figure~\ref{fig:DM_sens} shows the expected 95\% C.L. upper limits on the decay lifetime $\tau$ versus $M_{\rm DM}$ for DM particles decaying into $W^+ W^-$, $b\bar{b}$ and $\tau^+\tau^-$ assuming both an Einasto and a Burkert profile for the GC halo. Sensitivities are shown for 10 years of observations with SWGO. CTA sensitivity curves are also shown for 200 hours of observation of an Einasto profile, and decays into $b\bar{b}$ and $\tau^+\tau^-$~\cite{Pierre:2014tra}. The current most-stringent Fermi-LAT limits based on the observation and modeling of the isotropic gamma-ray background (IGRB)~\cite{Cohen:2016uyg} are also plotted. Current limits to DM decay in the GC halo have been provided by HAWC~\cite{Abeysekara:2017jxs}, and these are added to Fig.~\ref{fig:DM_sens} for comparison. 

\begin{figure}[h!]
	\begin{center}	
		\includegraphics[width=0.49\linewidth]{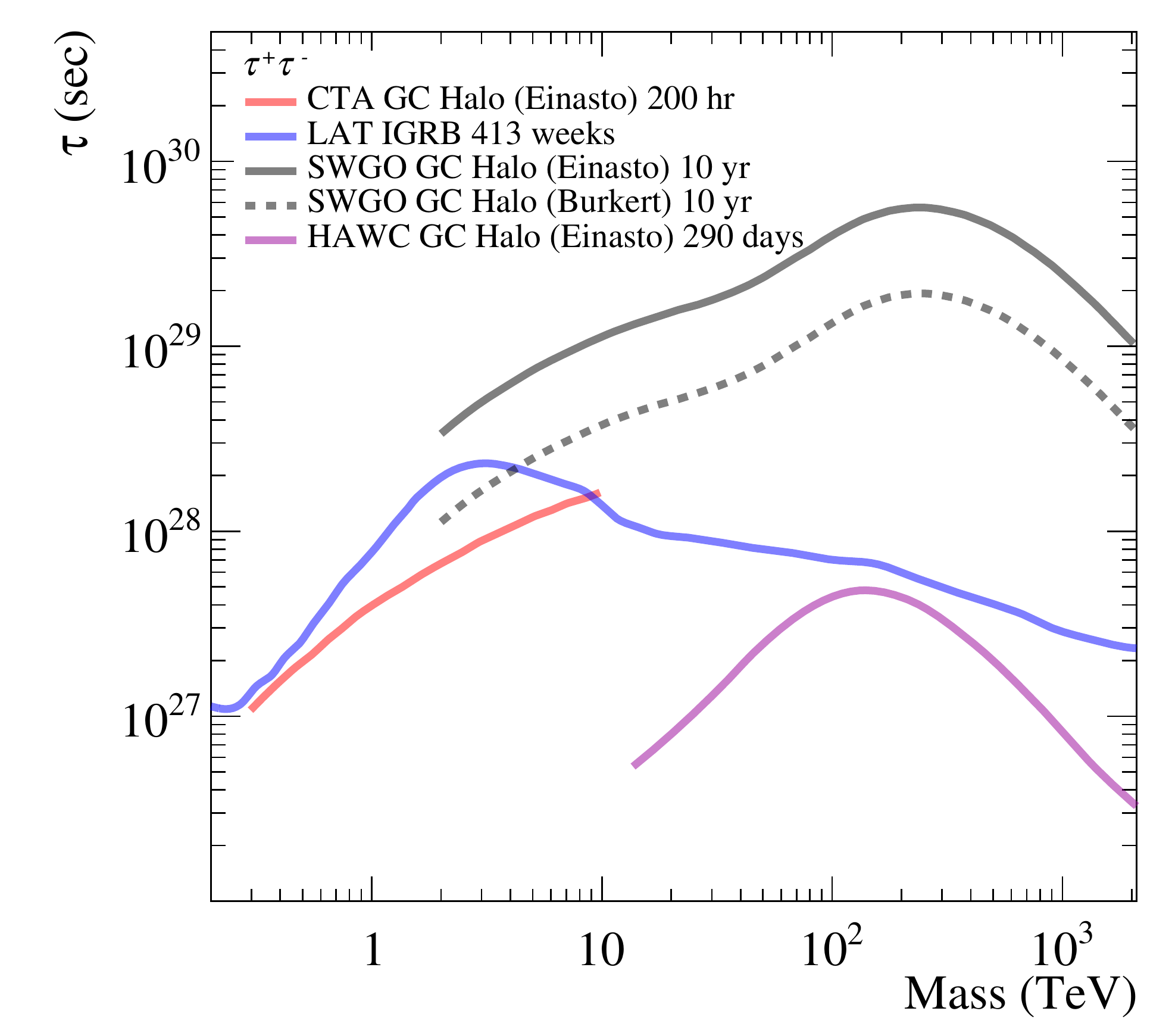}
		\includegraphics[width=0.49\linewidth]{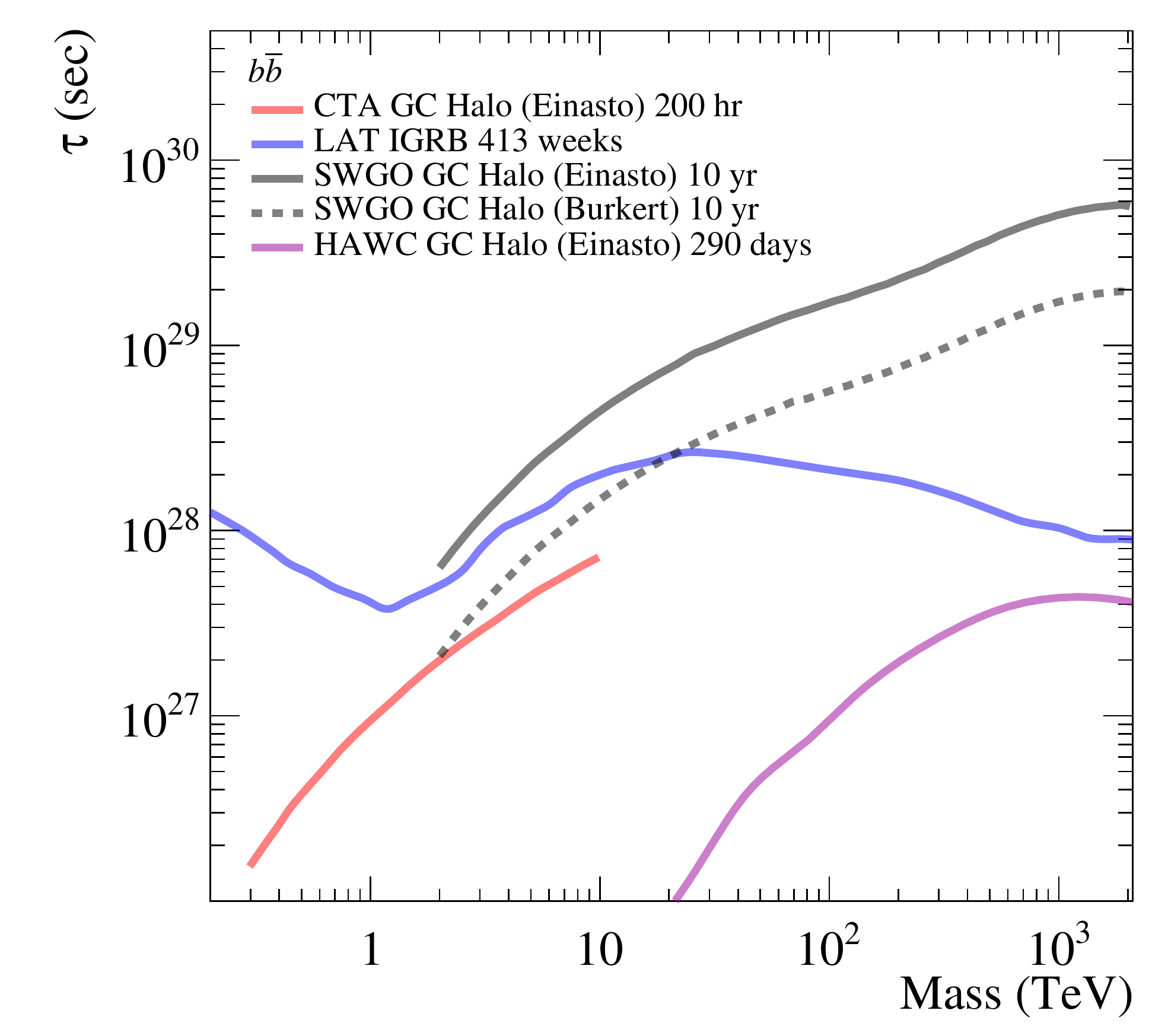}
		\includegraphics[width=0.49\linewidth]{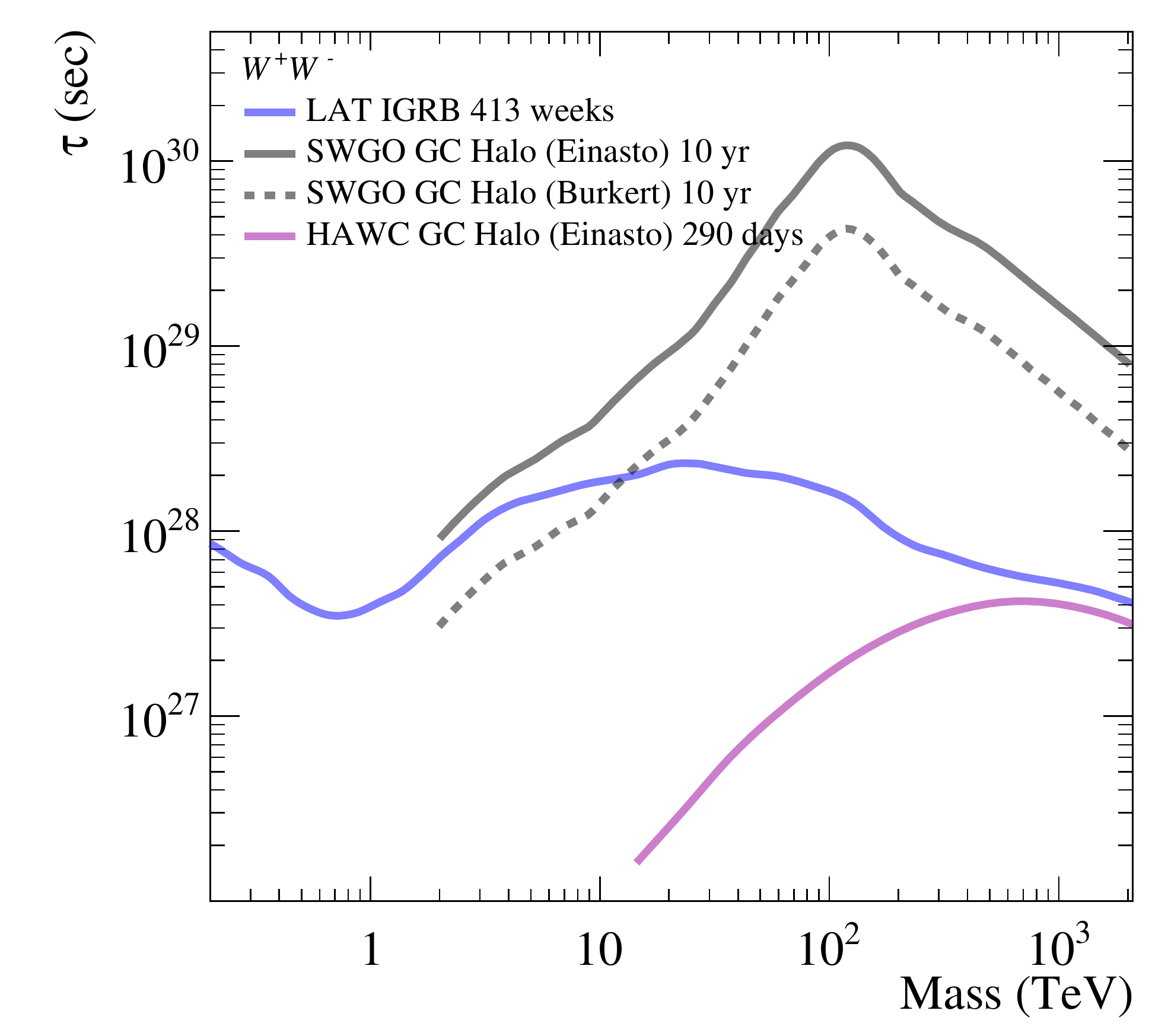}
	\end{center}
	\caption{Expected 95\% C.L. lower-limits on the DM decay lifetime into $\tau^+ \tau^-$ (top-left), $b\bar{b}$ (top-right) and $W^+ W^-$ (bottom) as a function of $M_{\rm DM}$, for SWGO observations of the GC halo, assuming both an Einasto (solid gray line) and a Burkert (dahsed gray line) profile. CTA sensitivity~\cite{Pierre:2014tra} and HAWC lower-limits~\cite{Abeysekara:2017jxs}  are also plotted for observations of an Einasto profile of the GC halo. Fermi-LAT~\cite{Cohen:2016uyg} lower-limits from observations of the isotropic gamma-ray background are shown as well.}
	\label{fig:DM_decay_sens}
\end{figure}

SWGO will reach an unprecedented sensitivity in the TeV mass range, being more sensitive than CTA and Fermi-LAT for all DM particle masses above $\sim$600 GeV. With respect to the current limits from HAWC~\cite{Abeysekara:2017jxs}, an improvement of more than two orders of magnitude is expected. A sensitivity to decaying lifetimes larger than 10$^{27}$ seconds will be attained for all channels and masses above 1 TeV.     

\subsection{Density profile effects}

\begin{figure}[h!]
	\includegraphics[width=0.85\linewidth]{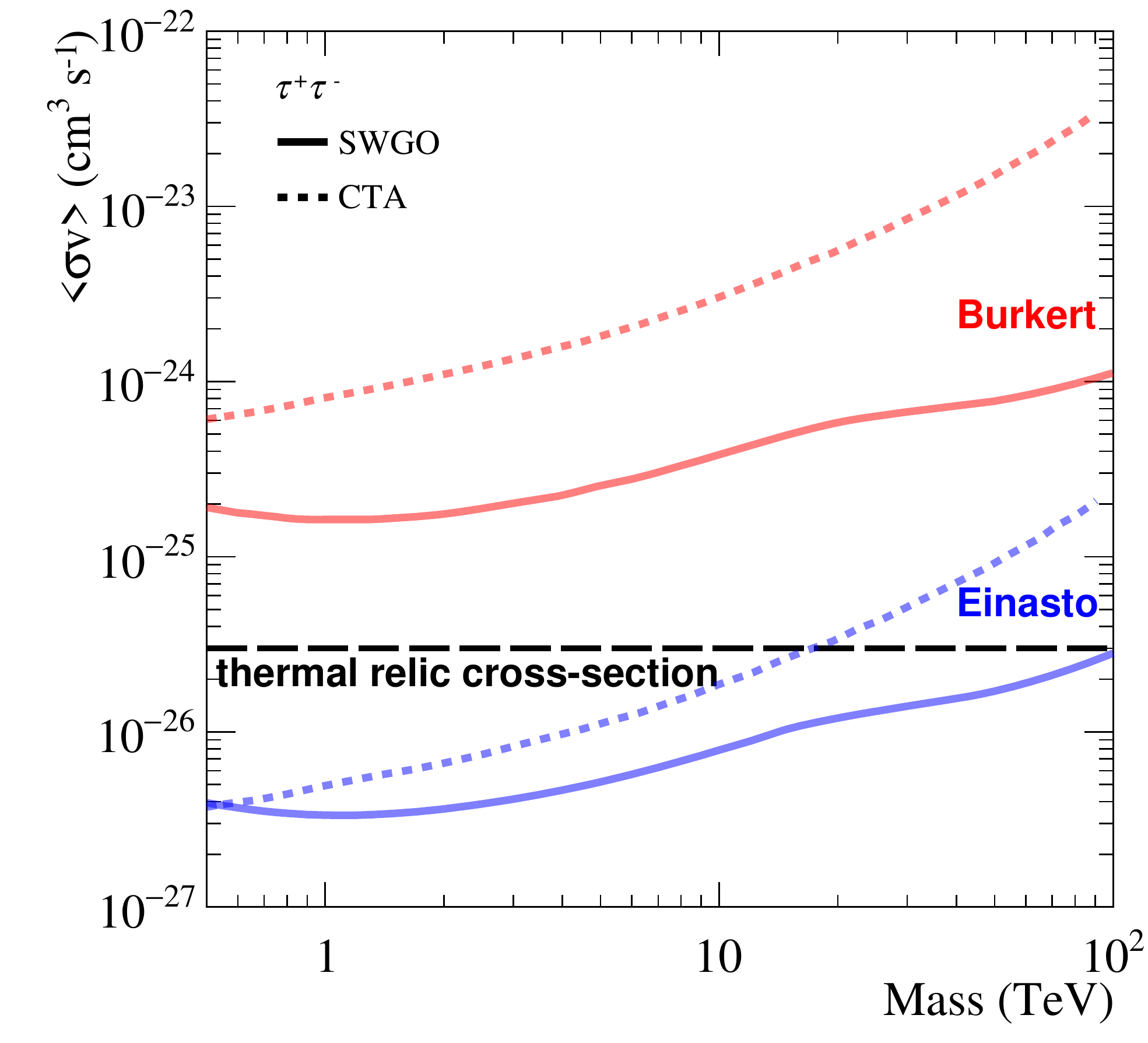}
	\caption{95\% C.L. sensitivity upper limit on the velocity-weighted cross section for DM annihilation into $\tau^+ \tau^-$ as a function of $M_{\rm DM}$, for both an Einasto (blue lines) and Burkert (red lines) profile of the Galactic halo. The sensitivity of SWGO (solid lines)is calculated in the inner 10$^{\circ}$, and CTA (dashed lines) in the inner 5$^{\circ}$ of the Galaxy~\cite{CTA_ScienceTDR,Lefranc:2016fgn}, excluding a $\pm$0.3$^{\circ}$ band in Galactic latitude. The nominal value of the thermal-relic cross-section is plotted as well (long-dashed black line).}
	\label{fig:DM_sens_comp}
\end{figure}

In order to estimate the impact of different Galactic halo profiles in the sensitivity estimates of different instruments we calculate the sensitivity assuming a Burkert profile and an Einasto profile. Figure.~\ref{fig:DM_sens_comp} shows the expected limits for annihilation into $\tau^+ \tau^-$ and Fig.~\ref{fig:DM_decay_sens} shows the expected limits for decay into all three channels. The sensitivity of CTA is also plotted for comparison~\cite{CTA_ScienceTDR,Pierre:2014tra}. Note that, although CTA observation strategy foresees a survey of the GC region, the relatively small FOV of CTA limits the size of the region that can be homogeneously covered. Hence, the signal extraction region of CTA was limited to the inner 5$^{\circ}$ of the Galaxy when searching for DM annihilation, and to the inner 1.29$^{\circ}$ for DM decay searches, although the latter may be too conservative. As already shown before, SWGO would be more sensitive to DM annihilation than CTA for all DM masses above 700 GeV assuming an Einasto profile, and this difference in sensitivity becomes even more pronounced for a cored Burkert profile. In this case, CTA limits would degrade by a factor of $\sim$166~\cite{Lefranc:2016fgn}, whereas SWGO limits would get weaker only by a factor of $\sim$48 (see Fig.~\ref{fig:DM_sens_comp}). The limits on WIMP annihilation are highly sensitive to the assumed behavior of the DM halo towards the center. If the DM density profile flattens toward the center, the expected flux from this region becomes much smaller and the limits become much less constraining. However, a survey-style instrument would be able to consider a more extended region surrounding the central halo and thus recover some of the integrated flux.   

Indeed, cored profiles are best observed using a wide FOV instrument. In addition to the increased sensitivity, a wide FOV allows for a robust estimate of the hadronic background.  Since the DM flux from a cored profile is roughly constant over space near the center,  backgrounds estimated from off regions too close to the region of interest would be highly contaminated by signal.  Wide FOV instruments are able to simultaneously observe regions far enough away from the halo center to eliminate signal contamination, allowing them to resolve emission even in the case of a cored profile.

An example of the power of simultaneous observation of background estimates for highly-extended sources is the TeV emission from the Geminga pulsar.  This emission has only been observed by wide FOV instruments such as Milagro~\cite{2009ApJ...700L.127A}, and the contemporary Fermi-LAT~\cite{DiMauro:2019yvh} and HAWC experiment~\cite{Geminga_HAWC}, while observations from IACTs such as the contemporary VERITAS experiment have shown no significant excess so far~\cite{Flinders:2015baa}.  




\subsection{Importance of electroweak corrections at TeV mass-scale}

Electroweak (EW) radiative corrections significantly modify the energy spectra of annihilation/decay of DM particles with masses larger than the electroweak scale~\cite{Ciafaloni:2010ti}. At energies much higher than the weak scale, the highly-energetic initial products of the DM annihilation/decay soft radiate electroweak gauge bosons W/Z, which then decay into many other SM particles. Most notably, the effect of these EW-corrections are particularly relevant for large DM masses (above a TeV).  

\begin{figure}[ht]
	\begin{center}
		\includegraphics[width=0.49\linewidth]{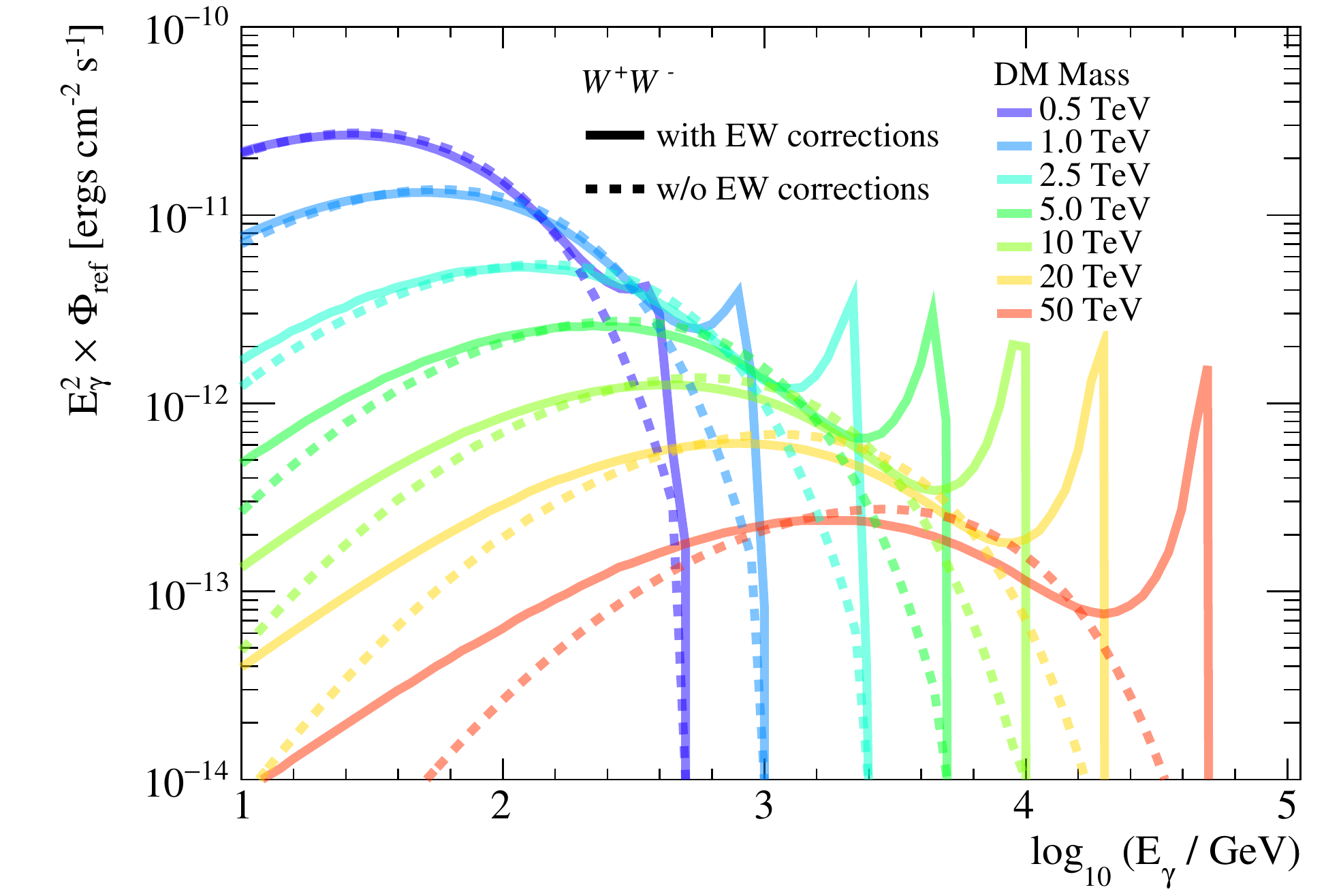}
		\includegraphics[width=0.46\linewidth]{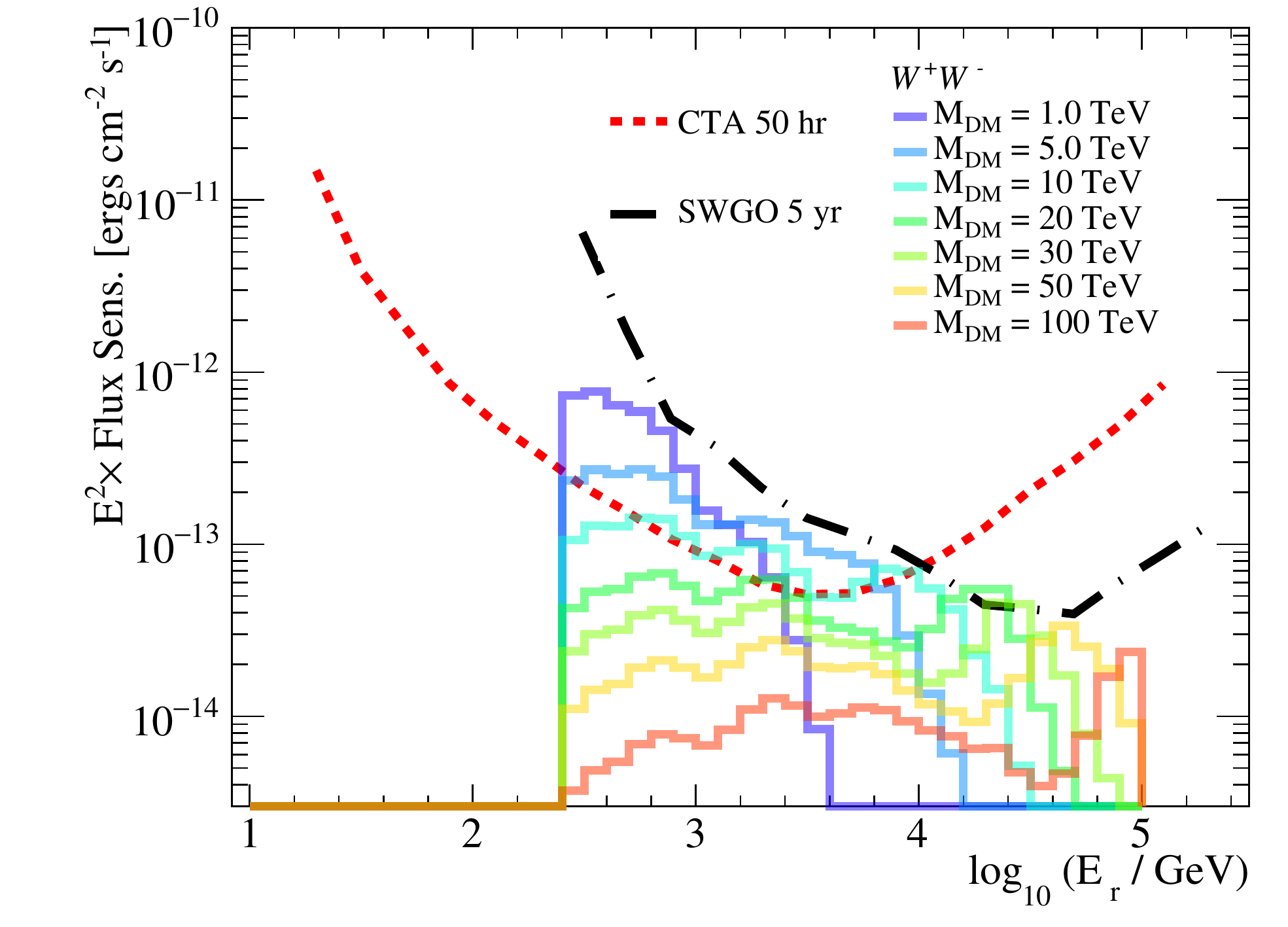}
	\end{center}
	\caption{\emph{Left:} DM annihilation spectra into $W^+ W^-$ with (solid lines) and without (dashed lines) taking into account electroweak corrections~\cite{Cirelli:2010xx} for different DM particle masses. Here, no convolution by the SWGO energy resolution was applied. \emph{Right:} SWGO and CTA flux sensitivity for point-like sources as a function of reconstructed gamma-ray energy, for 5 years and 50 hours of observation of the Galactic halo, respectively. Also plotted are the DM annihilation rate convolved by the SWGO energy resolution into $W^+ W^-$ per reconstructed energy bin (${\rm E_r}$) for different DM particle masses in arbitrary units, but keeping $\langle \sigma v\rangle$ and the $J$-factor the same for all masses.}
	\label{fig:DM_fluxes}
\end{figure}

\begin{figure}[h!]
	\begin{center}	
		\includegraphics[width=0.85\linewidth]{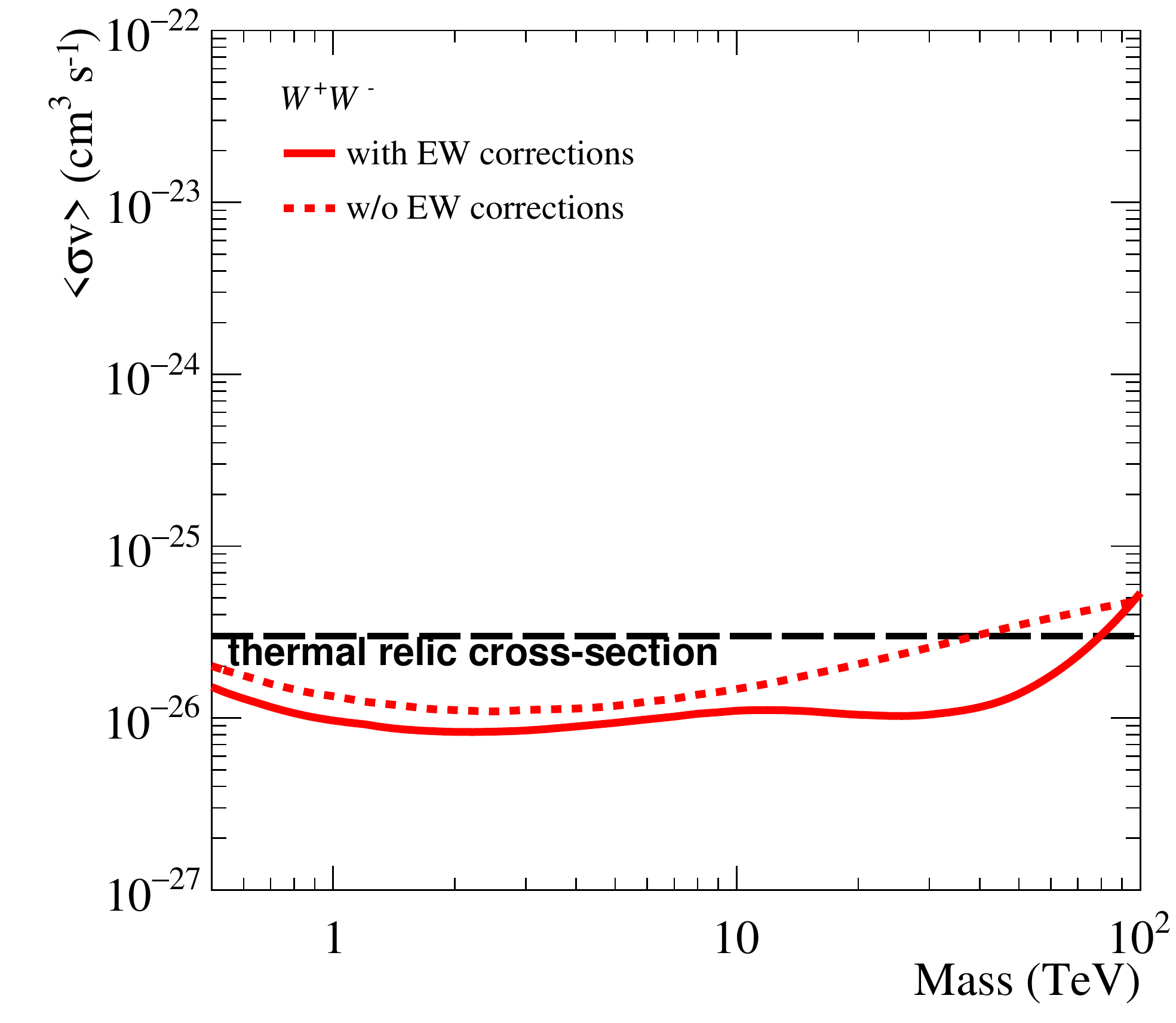}
	\end{center}
	\caption{Comparison of SWGO sensitivity upper limit for DM particles annihilating into $W^+ W^-$ with and without taking into account electroweak corrections. The nominal value of the thermal-relic cross-section is plotted as well (long-dashed black line).}
	\label{fig:DM_EW_corrections}
\end{figure}

 The typical modifications to the spectra of TeV-scale DM particles are: (\textit{i}) they enhance the low energy part of the spectrum, as a small number of highly-energetic particles are converted into a great number of low-energy particles; (\textit{ii}) since they open new channels in the final states which otherwise would be forbidden, all stable particles will be present in the final spectrum, independently of the primary annihilation channel considered; (\textit{iii}) in the case of an annihilation/decay into $W^+ W^-$, a strong peak close in energy to the value of the DM mass arises (see Fig.~\ref{fig:DM_fluxes}). In Fig.~\ref{fig:DM_EW_corrections}, we estimate the impact of such additional features for the SWGO sensitivity. For annihilation into $W^+ W^-$, EW-corrections improve the sensitivity by a factor of $\sim$1.3 for masses $\lesssim 10$ TeV. Between 20 and 80 TeV, the sensitivity gets better by as much as a factor of $\sim$2, creating a ``bump'' in the sensitivity curve in this mass range. This improvement comes mainly from the fact that SWGO reaches its deepest flux sensitivity exactly for energies where the peak in the DM particle spectra becomes more prominent, as it can be seen on the right panel of Fig.~\ref{fig:DM_fluxes}. 

\subsection{Complementarity between gamma-ray observatories}

The combination of deep observations of the GC region by SWGO with other gamma-ray observatories would provide independent results. In the case of a detection, it would increase confidence in the signal if it were found in both.
Observing the cutoff of the spectrum at the DM mass would be one of the strongest indications that an observed gamma-ray source originates from DM interactions and is the source hypothesis used to constrain DM interactions.  As shown in Fig. \ref{fig:DM_fluxes}, SWGO would achieve peak sensitivity at the energy scale where these cutoffs would be apparent for multi-TeV DM mass. In the case of a DM particle in the mass range 10 - 80 TeV annihilating into $W^+ W^-$, such an instrument would provide the DM mass measurement by probing the spectral cut-off, with CTA helping to constrain the morphology. We note that this mass range also has considerable advantages compared to the GeV range in terms of astrophysical foreground. There is a much shorter list of objects capable of accelerating particles to these energies and  avoiding the magnetospheric emission of pulsars, whose spectra can mimic an annihilation spectrum in the GeV range~\cite{Daylan:2014rsa,Hooper:2010mq}.

\section{Conclusion}

The inner Galactic halo is one of most promising regions for detecting gamma-ray signals from DM annihilation or decay.  Its close proximity and high-DM content yield one of the highest-expected fluxes from DM interactions. A survey-style instrument with a wide FOV in the Southern Hemisphere, such as SWGO, will be an important tool in searching for such emissions from multi-TeV DM mass. Here we propose a design of SWGO that would be sensitive enough to probe thermal DM for a large range of multi-TeV DM masses and interaction channels. In addition, the large FOV would allow for strong constraints on DM annihilation and decay even for density profiles that have large flat cores.  We also highlighted the impact of EW-corrections for the detection of DM particles with masses in the multi-TeV energy scale. A wide FOV experiment would also be able to work in tandem with CTA to confirm and identify any potential detection of DM emission through independent observations.  With all of these advantages available, a Southern wide FOV gamma-ray observatory promises to shed new light on the still unknown nature of DM. It will be a critical tool towards a better understanding of this diverse topic in the coming decade.

\section*{Acknowledgements}
AV and VdS have been supported by the S\~{a}o Paulo Research Foundation (FAPESP) through Grant No 2015/15897-1. AV work has been financed in part by the Coordena\c{c}\~{a}o de Aperfei\c{c}oamento de Pessoal de N\'{i}vel Superior - Brasil (CAPES) - Finance Code 001.


\printbibliography

\end{document}